\documentclass[article]{emulateapj}

\newcommand{\e}{\epsilon}
\newcommand{\g}{\gamma}

\newcommand{\ep}{\epsilon^\prime}

\newcommand{\psim}{\lower.5ex\hbox{$\; \buildrel \propto \over\sim \;$}}
\newcommand{\lbar}{\lower.0ex\hbox{$\; \buildrel
{\lower0.0ex \hbox{-}} \over\lambda  \;$}}

\shorttitle{Extragalactic Background Light}
\shortauthors{Finke, Razzaque, \& Dermer}

\begin{document}
\title {Modeling the Extragalactic Background Light from Stars and Dust}

\author{Justin D. Finke,$^{1}$ Soebur Razzaque$^{1}$ and 
	Charles D. Dermer}

\affil{U.S.\ Naval Research Laboratory, Code 7653, 4555 Overlook Ave SW,
        Washington, DC
        20375-5352\\
$^1$NRL/NRC Research Associate}

\email{justin.finke@nrl.navy.mil}

\begin{abstract}
The extragalactic background light (EBL) from the far infrared through
the visible and extending into the ultraviolet is thought to be
dominated by starlight, either through direct emission or through
absorption and reradiation by dust.  This is the most important energy
range for absorbing $\g$-rays from distant sources such as blazars and
gamma-ray bursts and producing electron-positron pairs.  In previous
work we presented EBL models in the optical through ultraviolet by
consistently taking into account the star formation rate (SFR),
initial mass function (IMF) and dust extinction, and treating stars on
the main sequence as blackbodies.  This technique is extended to
include post-main sequence stars and reprocessing of starlight by
dust.  In our simple model, the total energy absorbed by dust is
assumed to be re-emitted as three blackbodies in the infrared, one at
40 K representing warm, large dust grains, one at 70 K representing
hot, small dust grains, and one at 450 K representing polycyclic
aromatic hydrocarbons.  We find our best fit model combining the
Hopkins and Beacom SFR using the Cole et al. parameterization with the
Baldry and Glazebrook IMF agrees with available luminosity density
data at a variety of redshifts.  Our resulting EBL energy density is
quite close to the lower limits from galaxy counts and in some cases
below the lower limits, and agrees fairly well with other recent EBL
models shortward of about 5 $\mu$m.  Deabsorbing TeV $\g$-ray spectra
of various blazars with our EBL model gives results consistent with
simple shock acceleration theory.  We also find that the universe
should be optically thin to $\g$-rays with energies less than 20 GeV.
\end{abstract}

\keywords{galaxies: active --- 
	  diffuse radiation --- gamma rays:  observations --- 
	  stars:  luminosity function, mass function --- 
	  stars:  formation }

\section{Introduction}
\label{intro}

The extragalactic background light (EBL) from $\sim 10^{-3}$ -- 10 eV
($\sim 0.1$ -- $1000\,\mu$m) is thought to be dominated by starlight,
either through direct emission, or through absorption and reradiation
by dust.  This is the most important energy range for photons with
energy $E_1$(TeV) interacting with long-wavelength photons with and
absorbing $\g$-rays from distant sources, with the threshold
$\gamma\gamma$ condition implying that
$$
E_1(TeV) = \frac{0.26}{(1+z)\ E_{EBL}(eV)} \; .
$$
The EBL is also an important target radiation field for energy-loss
and/or photodisintegration of the highest energy cosmic ray protons
and ions \citep[e.g.,][]{puget76}, which results in a source of
high-energy neutrinos \citep[e.g.][]{stanev04}.  Direct measurement of
the EBL is difficult \citep[see][for a review]{hauser01} due to
contamination by foreground zodiacal and Galactic light.  Galaxy
counts may also be used to estimate the EBL, but the unknown number of
unresolved sources results in a lower limit.  The general picture is a
component peaking at around 1 $\mu$m from direct starlight emission
and one peaking at $\sim100$ $\mu$m from re-emission of absorbed
starlight by dust.  Due to different modeling approaches and
uncertainties in underlying model parameters, the intensity and shape
of the EBL spectrum remains controversial.

A wide range of EBL models have been developed.  One class of models
approaches the problem by using IR data from local galaxies and
extrapolating their evolution to higher redshifts and shorter
wavelengths \citep[e.g.,][]{malkan98,malkan01,stecker06}.  Another
class of models uses semi-analytic merger-tree models of galaxy
formation to determine the star formation history of the universe for
a forward-evolution calculation
\citep{primack01,primack05,primack08,gilmore08,gilmore09}.  The class
of models considered here focuses on the primary energy source of the
emission, namely the stars. In such models, one integrates over star
formation rates and stellar properties
\citep[e.g.][]{salamon98,dwek98_theory,kneiske02,kneiske04}. Alternatively,
luminosity density data \citep{franceschini08} is used to estimate the
EBL energy density and its evolution with time.  Since the EBL absorbs
very high energy (VHE; $\approx 0.1$ -- 100 TeV) $\g$-rays from
blazars, one can use this mechanism to put upper limits on the EBL
energy density
\citep[e.g.,][]{stecker93,stanev98,aharonian06,mazin07,finke09}.
Disagreement about the intrinsic $\g$-ray spectra of blazars
\citep[e.g.,][]{stecker07_accel,boett08} has resulted, however, in a
lack of consensus about the meaning of these upper limits.  Detection
of a 33 GeV photon from GRB 090902B \citep{abdo09_090902b} and a 13
GeV photon from GRB 080916C \citep{abdo09_080916c} with the Large Area
Telescope (LAT) onboard the {\em Fermi} Gamma Ray Space Telescope are
used to test different EBL models, which may be a promising new way to
constrain the EBL.

We \citep*[][hereafter RDF09]{razzaque09} recently developed models
for starlight luminosity density and EBL intensity by assuming that
stars are blackbodies, and integrating over stellar properties on the
main sequence.  Here we extend the model of RDF09 to stars that have
evolved off the main sequence and include re-emission of absorbed
starlight by dust, noting that the models of RDF09 remain valid for
the $\ga 1$ eV range. In \S\ \ref{lightmodel}, analytic expressions
for the radiation from stars and re-emission by dust are presented.
Numerical calculations of the luminosity density and EBL energy
density are presented in \S\ \ref{numericalresults}.  We explore the
effects of our EBL model on absorption of distant $\g$-rays (\S\
\ref{absorption_sec}) and conclude with a discussion on our results
and future research (\S\ \ref{summary}).

\section{Formalism}
\label{lightmodel}

We briefly describe the RDF09 model for background starlight and our
recent improvements which include the emission from post-main sequence
stars and dust.  Integrating over star formation in this manner is
similar to several other models
\citep[e.g.][]{salamon98,dwek98_theory,kneiske02,kneiske04}.

\subsection{Direct Starlight Emission}
\label{starlight}

Stars with dimensionless mass $m = M/M_{\odot}$ and age $t_{\star}$ are
assumed to emit as blackbodies.  The photon density of a 
blackbody is given by
\begin{equation}
\label{blackbody}
n_{\star}(\e;m,\Theta) = \frac{dN}{d\e dV} = 
	\frac{8\pi}{\lambda_C^3}\ \frac{\e^2}{\exp[\e/\Theta] - 1}
\end{equation}
where $\e=h\nu/m_ec^2$ is the dimensionless photon energy,
$\lambda_C=h/m_ec\approx 2.42\times10^{-10}$ cm is the Compton
wavelength, and $\Theta=k_BT/m_ec^2$ is the dimensionless effective
temperature.  The total number of photons emitted per unit
energy per unit time from a star of radius $R(m,t_\star)$ and
effective stellar temperature $\Theta(m,t_\star)$ is 
\begin{equation}
\dot{N}_{\star}(\e; m,t_\star) = \frac{dN}{d\e dt} = \pi R(m,t_\star)^2c\ 
	n_{\star}(\e;\Theta(m,t_\star)\ .
\end{equation}

To determine the luminosities and radii of the stars, $L(m,t_\star)$
and $R(m,t_\star)$, respectively, as well as the time stars spend on
various portions of the Hertzsprung-Russell diagram, we used the
stellar formulae from the appendix of \citet{eggleton89}.  These
formulae approximate the stellar parameters along the main sequence,
the Hertzsprung gap, the giant branch, the horizontal branch, the
asymptotic giant branch, and the white dwarf phase for stars of solar
metalicity.  Thus, we assume all stars emitting since star formation
began have solar metalicity.  Note that eqn.\ (A15) of
\citet{eggleton89}, which describes the luminosity of the base of the
giant branch, should be
$$
L_{BGB} = \frac{ 2.15M^2 + 0.22 M^5 } 
               {1 + 1.4\times 10^{-2}M^2 + 5\times 10^{-6}M^4 }
$$
\citep{eggleton90}.  Also, eqn.\ (A22) of \citet{eggleton89}, 
which describes the time a star spends burning Helium, should be
$$
t_{He} = \frac{ t_{MS}L_0 }
              { L_{He} (M^{0.42} + 0.8) }
$$
(C. Tout and P. Eggleton, private communication).  
We modify eqn.\ (A27) 
of \citet{eggleton89}, which describes the luminosity of a white dwarf 
in our calculations, so that it reads
$$
L = \frac{40}{(t - t_{WD} + 0.1)^{1.4}}
$$
in order to avoid a singularity when $t = t_{WD}$ \citep{hurley00}.
In the above corrections to \citet{eggleton89}, we use their
notation, so that $M$ is the star's mass in units of $M_{\odot}$,
$L_{BGB}$ and $L_0$ are in units of $L_{\odot}$, and $t$, $t_{He}$,
$t_{MS}$, and $t_{WD}$ are in units of Myr.

Once a star's luminosity and radius have been
determined, its temperature can be found by
\begin{equation}
\Theta(m,t_\star) = \frac{ k_BT_\odot }{m_ec^2} 
	\left( \frac{L(m,t_\star)}{L_\odot} \right)^{1/4}
	\sqrt{ \frac{R_\odot}{R(m,t_\star)} }\ 
\end{equation}
where $T_\odot=5777$ K is the effective solar temperature,
$L_\odot=3.846\times10^{33}$ erg s$^{-1}$ is the solar luminosity, and
$R_\odot=6.96\times10^{10}$ cm is the solar radius.

We will use unprimed symbols to refer to quantities measured in the
frame comoving with the Hubble flow, and primed or double primed
quantities to refer to quantities in the proper frame of a galaxy at
some redshift $z$.  Observations made from the solar system,
which has a small peculiar velocity with respect to the Hubble flow,
are essentially in the comoving frame.  The comoving luminosity
density (i.e., the luminosity per unit comoving volume, or the
emissivity) of the universe as a function of comoving energy $\e$ at a
certain redshift $z$ (in units of, e.g., W Mpc$^{-3}$) can be found
from
\begin{eqnarray}
\label{lumdens}
\e\, j^{stars}(\e; z) = m_ec^2 \epsilon^{2}\,
	\frac{ dN}{dt\, d\e\, dV} = 
\nonumber \\ 
	m_ec^2 \epsilon^{2} f_{esc}(\e) \int^{m_{max}}_{m_{min}} dm\ 
	\xi(m)\ \  
\nonumber \\ \times
	\int^{z_{max}}_{z} dz_1\ \left|\frac{dt_*}{dz_1}\right|\ 
	\psi(z_1)\ \dot{N}_{\star}(\e; m,t_\star(z,z_1))\ .
\end{eqnarray}
The luminosity density is dependent on the initial mass function
(IMF), $\xi(m)$, the comoving star formation rate (SFR) density (i.e.,
the star formation rate per unit comoving volume), $\psi(z)$, and the
fraction of photons which escape a galaxy, $f_{esc}(\e)$ and avoid
being absorbed by interstellar dust.  The relationship between cosmic
time and redshift is given by
\begin{eqnarray}
\label{dtdz}
\left| \frac{dt_*}{dz}\right|\ = \frac{1}
{H_0(1+z)\sqrt{\Omega_m(1+z)^3 + \Omega_\Lambda}}\ ,
\end{eqnarray}
in a flat $\Lambda$CDM cosmology.  
We assume cosmological parameters $H_0=70$ km s$^{-1}$ Mpc$^{-3}$, 
$\Omega_m=0.3$, and $\Omega_\Lambda=0.7$.  

\citet{driver08} have applied the dust model of \citet{popescu09} to a
survey of $\sim10^5$ nearby galaxies from the Millennium Galaxy
Catalog \citep{allen06} to determine the wavelength-dependent escape
fraction of photons in the local universe.  RDF09 have fit this with a
series of power-laws and we use this to compute $f_{esc}(\e)$.  We
also assume that any photon with $m_ec^2\e>13.6$ eV is absorbed by
galactic and intergalactic \ion{H}{1} gas. These UV photons are not
reprocessed in our model, and we assume their net energy makes a
small contribution to the total EBL intensity.

In this work we choose the limits of integration $m_{min}=0.1$,
$m_{max}=100$, and $z_{max}=6$, although our EBL intensities and
luminosity density results at low $z$ are not strongly dependent on
the upper limits.  Thus, the model of the stellar component does not
have any adjustable parameters once an IMF and SFR have been chosen.

\begin{figure}
\epsscale{1.0}
\plotone{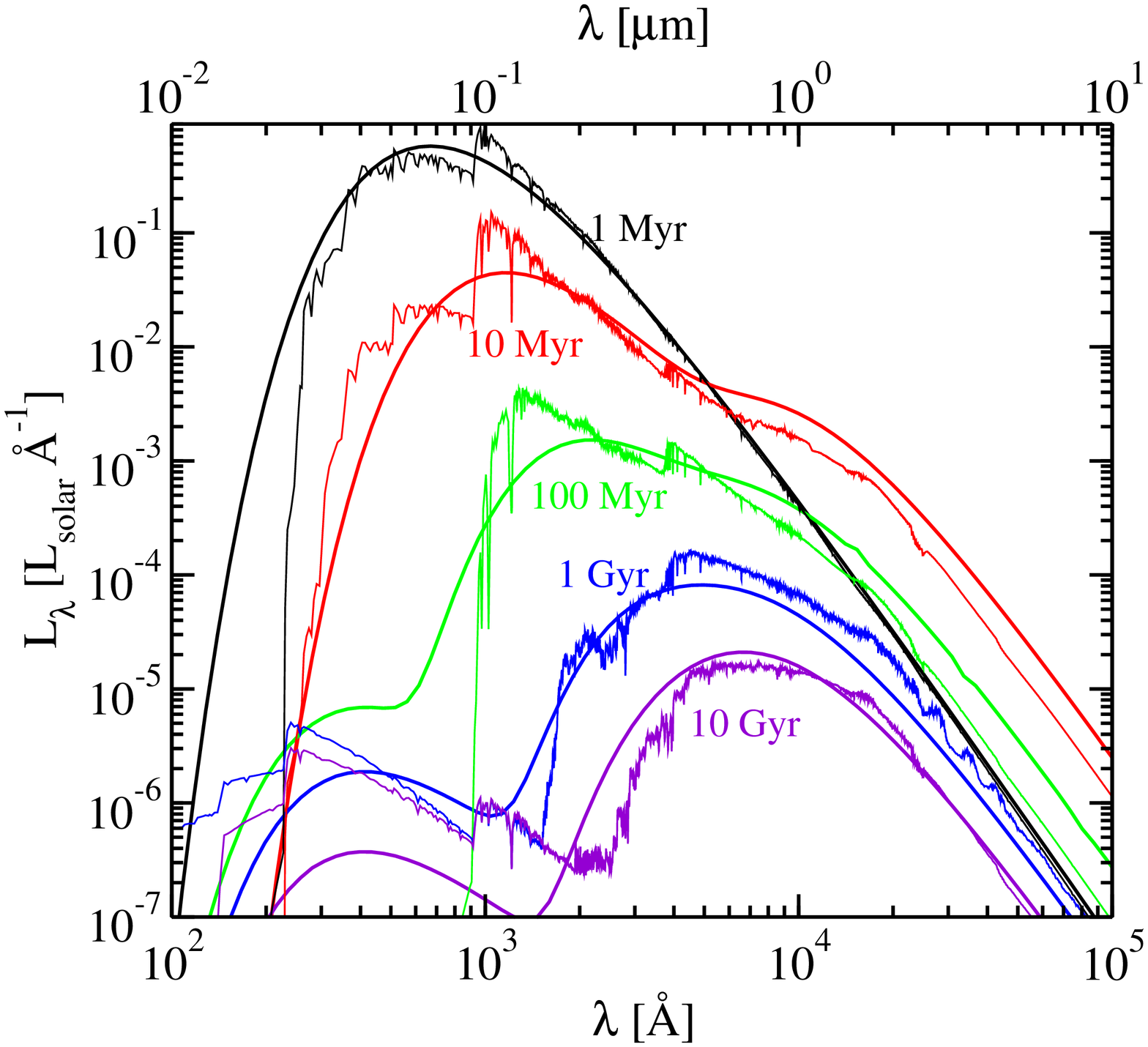}
\caption{
SSP spectra for various population ages, for a Salpeter IMF,
normalized to 1 $M_\odot$.  The thin lines are the models from
\citet{bruzual03} and the thick lines are our calculations assuming
stars are blackbodies. Photons with wavelengths shortward of 912 \AA\
(13.6 eV) contribute a negligible amount to the total energy budget
and are assumed to be completely absorbed.
}
\label{ssp_fig}
\end{figure}

To test the accuracy of approximating stars as blackbodies and the
simple \citet{eggleton89} stellar formulae, we computed the spectra of
simple stellar populations (SSPs) for various ages from
\begin{eqnarray}
\label{ssp_formula}
L_\lambda(t_\star) = m_ec^2\e\, \frac{dN}{dt\,d\lambda} = \
\nonumber \\ 
\frac{m_e^2c^4\e^3 }{hc}
	\int^{m_{max}}_{m_{min}} dm\ \xi(m)\ 
	\dot{N}_{\star}(\e; m,t_\star)\ 
\end{eqnarray}
using a \citet{salpeter55} IMF, i.e., $\xi(m) \propto m^{-2.35}$, from
$m_{min}=0.1 < m < m_{max}=100$.  Eqn. (\ref{ssp_formula})
approximates the spectrum of a cluster of stars that were all born at
exactly the same time, and all now have an age of $t_\star$.  SSPs at
high spectral resolution were calculated by \citet{bruzual03} using
various stellar spectral libraries calculated with detailed stellar
structure codes.  The \citet{bruzual03} SSPs with solar metalicity are
compared with our approximations in Fig.\ \ref{ssp_fig}.  Our
approximations agree fairly well longward of $\sim10^3$ \AA\ (0.1
$\mu$m), particularly for stars between 1 and 10 Gyrs of age.  This
age range of stars when weighted by the SFR dominates the radiative
output between 0.1 and 10 $\mu$m, and this is where our approximations
are best.  Shortward of this wavelength the agreement is poor, as
large as several orders of magnitude.  However, at $\la 0.1\ \mu$m
most photons will be absorbed by \ion{H}{1} gas at low $z$, and at
high $z$ where stellar populations are younger agreement is much
better at these wavelengths, differing by no more than a factor
of $\sim$2--3.  We have computed a test EBL model using a Salpeter IMF
for both the blackbody approximation and using the \citet{bruzual03}
SSPs and found that the difference is no more than $\approx 25$\%.  As
we show below, this is larger than the difference without including
post-main sequence stars (\S\ \ref{sec_compare}).  We therefore
consider our blackbody approximation quite good for calculating
luminosity densities and EBL energy densities, particularly compared
to other uncertainties in these calculations.

%\clearpage

%We also consider a model based on \citet{fardal07} which uses a
%``diet'' Salpeter IMF and their own SFR. The diet
%Salpeter SFR is identical to the regular Salpeter SFR except that it
%extends down only to $m=0.788$, and the normalization is adjusted 
%accordingly.  The Fardal SFR is parameterized as
%\begin{equation}
%\label{fardalSFR}
%\psi(z) = \frac { p_1p_2p_3\rho_0(1+z)^{p_2} }
%	        { \left[ 1 + p_1(1+z)^{p_2}\right]^{p_3+1} }\ 
%	H(z)
%\end{equation}
%where 
%\begin{equation}
%H(z) = H_0 (1+z) \sqrt{ \Omega_m(1+z)^3 + \Omega_\Lambda }\ .
%\end{equation}
%They fit this parameterization to a large number of SFR data to obtain 
%best-fit values of $\rho_0=9.0\times10^{8} M_\odot$ Mpc$^{-3}$, 
%$p_1=0.075$, $p_2=3.7$, and $p_3=0.84$.  Thus, we also have:
%\itemize
%\item{ {\em Model F}:  \citet{fardal07} SFR and diet Salpeter IMF

\subsection{Dust Emission}
\label{dustsection}

The starlight that is absorbed by dust is reradiated in the infrared.
There are generally thought to be three major dust components in the
interstellar medium \citep[e.g.][]{desert90}: (1) a large grain
component that absorbs optical light and reradiates in the far-IR,
found in and around star-forming regions; (2) A small grain component
that absorbs the far UV and reradiates in the near-IR, located
throughout the disk of spiral galaxies and is responsible for most of
the observed dust radiation and (3) a component from polycyclic
aromatic hydrocarbons (PAHs) which emit as broad emission lines and
are not generally in thermodynamic equilibrium with their environment
\citep{dwek97}, although they are sometimes modeled as very hot
blackbodies \citep[e.g.][]{kneiske02}.

We assume the dust emits as a combination of three blackbodies,
representing three types of dust: a 40 K blackbody representing warm,
large dust grains, a 70 K blackbody representing hot, small dust
grains, and a 450 K blackbody representing PAHs.  These
temperatures are similar to those used by \citet{kneiske02}.  By
setting the luminosity density from dust emission equal to the
luminosity density from starlight absorbed from dust, we calculate the
infrared emission self-consistently.  The emissivity absorbed by dust
is
$$
\int d\e\, \frac{1}{f_{esc}(\e)}\, [1-f_{esc}(\e)]\ j^{stars}(\e; z) 
$$
where $j^{stars}(\e; z)$ is the stellar emissivity (\S\
\ref{starlight}).  A fraction of this will be reradiated in each
component, so that
\begin{eqnarray}
\label{dustequiv}
f_n \int d\e\, \frac{1}{f_{esc}(\e)}[1-f_{esc}(\e)]\ 
	j^{stars}(\e; z) \ =
\\ \nonumber
	\int d\e\, j_{n}(\e; \Theta_n)\ 
\end{eqnarray}
where $f_n$ is the fraction of the absorbed emissivity reradiated in a
particular dust component, $\Theta_n = k_{\rm B}T_n/m_ec^2$ is the dimensionless temperature
of the dust component, and the subscripts $n=$1, 2, and 3 refer to the
warm dust, hot dust, and PAHs, respectively.  Since the dust radiates
as a blackbody, i.e.,
\begin{eqnarray}
j_{n}(\e; \Theta_n) = j^0_{n}
	\frac{\e^{3}}{\exp[{\e/\Theta_n}] - 1}\ ,
\end{eqnarray}
the integral in the right side of eqn.\ (\ref{dustequiv}) can
readily be performed, so that
\begin{equation}
\int_0^\infty d\e\, j_{n}(\e; \Theta_n) = 
	\frac{\pi^4}{15} j^0_{n} \Theta_n^4\ .
\end{equation}
This allows one to determine the blackbody normalization, 
\begin{eqnarray}
j^0_{n} = 15 \frac{ f_n \int d\e\ \left[\frac{1}{f_{esc}(\e)}-1\right]\ 
	j^{stars}(\e; z) }
	{ \pi^4\Theta_n^4 }\ .
\end{eqnarray}
Hence, the comoving luminosity density due to dust is
\begin{eqnarray}
\label{dust_lumdens}
\e\, j^{dust}(\e; z) = 
\frac{ 15 }{\pi^4 }  
\int d\e\ \left[ \frac{1}{f_{esc}(\e)}-1 \right]\ 
\nonumber \\ \times
	j^{stars}(\e; z) \,
\sum_{n=1}^3 \frac{f_n}{\Theta_n^4}\ \frac{\e^{ 4}}
	{\exp[\e/\Theta_n] - 1}\ .
\end{eqnarray}
We assume that $f_{esc}$, $f_n$, and $\Theta_n$ do not vary with $z$.
The values for $f_n$ and $\Theta_n$ were chosen by trial and
error to fit the IR luminosity density data at $z=0.0$ and $z=0.1$
(see \S\ \ref{lumdens_section}).  These values can be found in
Table \ref{dusttable}.  With the stellar component of the luminosity
density calculated from eqn. (\ref{lumdens}), the dust component of
the luminosity density can be determined.  Both the stellar and dust
luminosity densities are used to model the EBL energy density.

\subsection{Extragalactic Background Light Model}

The comoving EBL energy density at the current epoch can be obtained
by integrating the luminosity density, eqns.\ (\ref{lumdens}) 
and (\ref{dust_lumdens}),
\begin{eqnarray}
\e\, u_{EBL}(\e;z=0) = 
	m_ec^2 \e^{ 2}\, \frac{dN}{d\e\, dV} = 
\nonumber \\ 
	\ \int^{z_{max}}_{0}dz_1\ 
	\frac{ \ep\, j(\ep; z_1)}{(1+z_1)}\ 
	\left|\frac{dt_*}{dz_1}\right|\ 
\end{eqnarray}
where $\e\,j(\e; z) = \e\, j^{stars}(\e; z) + \e\, j^{dust}(\e; z)$
and $\ep=(1+z_1)\e$.  If star formation and star emission ended at
some redshift $z>0$, then the EBL energy density observed today 
would be 
\begin{eqnarray}
\e\, u_{EBL}(\e;z) = 
	\ \int^{z_{max}}_{z}dz_1\ 
	\frac{ \e^{\prime} j(\e^{\prime}; z_1)}{(1+z_1)}\ 
	\left|\frac{dt_*}{dz_1}\right|\ .
\end{eqnarray}
This is the energy in the comoving frame per unit comoving volume
as a function of $\e$, which is the photon energy in the comoving
frame.  The energy and volume in the comoving frame are quantities in
coordinates which expand with the universe, and hence are the
quantities which we observe today.  

However, observers in a galaxy at a redshift $z>0$ would have viewed
a universe that is $(1+z)^3$ smaller than now, and photons observed 
now with energy $\e$ would be observed by them to have 
proper energy $\e_p = (1+z)\e$.  The proper energy density they 
would have observed is given by
\begin{eqnarray}
\label{n_EBL}
\e_{p}\ u_{EBL,p}(\e_{p};z) =
	(1+z)^4\ \e\ u_{EBL}(\e;z) 
\nonumber \\ 
	= (1+z)^4\ \int^{z_{max}}_{z}dz_1\ 
	\frac{ \e^{\prime} j(\e^{\prime}; z_1)}{(1+z_1)}\ 
	\left|\frac{dt_*}{dz_1}\right|\ 
\end{eqnarray}
where
$$
\e^{\prime} = \e(1+z_1) = \frac{1+z_1}{1+z}\e_p\ .
$$
This is the radiation field with which a
$\g$-ray photon emitted at redshift $z$ interacts on its way to
Earth.  

Energy density can be converted to intensity in units of,
e.g., nW m$^{-2}$ sr$^{-1}$ by
\begin{eqnarray}
\e\, I(\e; z) = \frac{c}{4\pi}\ \e\, u_{EBL}(\e;z)\ ,
\end{eqnarray}
which is convenient for comparing to EBL intensity measurements.

%\clearpage
\begin{deluxetable}{lcccc}
%\rotate
\tabletypesize{\scriptsize}
\tablecaption{
Dust Parameters.  See text for details.
}
\tablewidth{0pt}
\tablehead{
\colhead{ Component} &
\colhead{ $n$ } &
\colhead{ $f_n$ } &
\colhead{ $T_n$ [K]} &
\colhead{ $\Theta_n$ [$10^{-9}$] }
}
\startdata
Warm Large Grains & 1 & 0.60 & 40 & 7\\
Hot Small Grains & 2 & 0.05 & 70 & 12 \\
PAHs & 3 & 0.35 & 450 & 76 \\
\enddata
\label{dusttable}
\end{deluxetable}
%\clearpage

\section{Numerical Results}
\label{numericalresults}

In this section, we compute the luminosity densities and EBL energy
densities from starlight and dust for various combinations of SFRs and
IMFs.  Although the \citet{salpeter55} IMF ($\xi(m) \propto
m^{-2.35}$) is still in common use, it now seems that it overpredicts
the number of low mass stars.
\citet{baldry03} fit local luminosity density data to find a preferred
IMF, which is similar to a Salpeter IMF above $m=0.5$ ($\xi(m) \propto
m^{-2.2}$) but flatter below this mass ($\xi(m) \propto m^{-1.5}$).
The Salpeter A IMF is also often used, which is $\xi(m) \propto
m^{-0.5}$ below $m=0.5$ and $\xi(m) \propto m^{-2.35}$ above $m=0.5$.
\citet{hopkins06} compiled a large amount of SFR data and fit them
using various IMFs.  To parameterize their SFR fits they used the
function of \citet{cole01} as well as their own piecewise power-law
function, and found the Salpeter A and \citet{baldry03} IMFs were
favored, under the assumption of a universal IMF.

For our models we use the same designations as RDF09 which are
as follows (see RDF09 for details on these SFRs and IMFs):
\begin{itemize}
\item {\em Model A}: \citet{cole01} SFR and Salpeter A IMF
\item {\em Model B}: \citet{hopkins06} SFR using \citet{cole01} 
	parameterization and Salpeter A IMF
\item {\em Model C}: \citet{hopkins06} SFR using \citet{cole01} 
	parameterization and \citet{baldry03} IMF.
\item {\em Model D}: \citet{hopkins06} SFR and Salpeter A IMF
\item {\em Model E}: \citet{hopkins06} SFR and \citet{baldry03} IMF
\end{itemize}
Measurements of SFRs are generally only sensitive to the formation of
massive stars, so that an IMF must be chosen to extrapolate to lower
masses and determine the global SFR.  Thus, the SFR and IMF are not
independent of one another and a combination of both are needed to
calculate luminosity densities and EBL energy densities.

\subsection{Luminosity Density}
\label{lumdens_section}

The luminosity density at a particular wavelength is measured by
adding up the luminosities of the galaxies at that wavelength in a
particular volume, and forming a luminosity function using the
$1/V_{max}$ method \citep[e.g.,][]{sandage79}.  This involves
finding the luminosities within a redshift slice, which gives the
maximum comoving volume ($V_{max}$) within which the galaxies could be
found.  Since the surveys are limited to a minimum observable
magnitude, some extrapolation to lower luminosities must be done,
which involves a fit to the luminosity function.  In the optical and
UV, the luminosity function is typically fit with a
\citet{schechter76} function which is
\begin{equation}
\phi(L) = \frac{\phi_*\ (L/L_*)^{\alpha}\ \exp(-L/L_*)}{L_*}\ ,
\end{equation}
where the function has three fit parameters, $\phi_{*}$, $L_*$, and
$\alpha$.  This function can be integrated to give the luminosity
density, so that
\begin{equation}
j_{\e} = \int^{\infty}_{0}\ dL\ L\ \phi(L) 
	= \phi_*L_*\Gamma(\alpha+2)\ 
\end{equation}
where $\Gamma(x)$ is the Gamma function.
The Schechter function, written in terms of absolute magnitude, $M$ 
(not to be confused with stellar mass in \S\ \ref{lightmodel}), is 
\begin{eqnarray}
\phi(M) = 0.4 \ln(10) \phi_{*}10^{0.4(M-M_{*})(\alpha+1)}
\\ \nonumber \times
	\exp (-10^{-0.4(M-M_{*})} )\ ,
\end{eqnarray}
which can then be integrated to give
the luminosity density in terms of AB magnitudes per volume,
\begin{equation}
j_M = M_* - 2.5\log\left[\phi_*\Gamma(\alpha+2)\right] + C
\end{equation}
where $C$ is the conversion from observed magnitudes to AB magnitudes.
The error in the luminosity density is
\begin{eqnarray}
\sigma_{j_M}^2 =
	\left( \frac{\partial j}{\partial M_*}\sigma_{M_*} \right)^2 + 
	\left( \frac{\partial j}{\partial \phi_*}\sigma_{\phi_*} \right)^2 +
	\left( \frac{\partial j}{\partial \alpha}\sigma_{\alpha} \right)^2 
\\ \nonumber
	= 
	\sigma_{M_*}^2 + 
	\left( \frac{2.5}{\ln(10)}
	\frac{\Gamma(\alpha+2)\ \sigma_{\phi_*}}{\phi_*} \right)^2 + 
\\ \nonumber
	\left( \frac{2.5}{\ln(10)} 
	\psi_0(\alpha+2)\ \sigma_\alpha \right)^2\ ,
\end{eqnarray}
where $\psi_0(x)$ is the digamma function (not to be confused with 
$\psi(z)$, the SFR, used in \S\ \ref{starlight}).  This can be converted 
to power per volume (in, e.g., W Mpc$^{-3}$) by 
\begin{eqnarray}
\e j_{\e} = \nu\ \times\ 10^{ \frac{j_M-34.10}{-2.5} }
\end{eqnarray}
and
\begin{equation}
\sigma_{\e j_{\e}} = \frac{\nu \ln(10)}{-2.5}\ 
		10^{ \frac{j_M-34.10}{-2.5} }\ \sigma_{j_M}
\end{equation}
where $\nu$ is the central frequency of the bandpass.

Luminosity functions in the IR can be quite different from the
optical, and are often fit with broken power-laws,
\begin{equation}
\phi(L) =  \left\{ \begin{array}{ll}
	   \phi_*\ \left(\frac{L}{L_*}\right)^{1-\alpha} 
	   & L<L_* \\
	   \phi_*\ \left(\frac{L}{L_*}\right)^{1-\beta}
	   & L>L_* 
	   \end{array} 
	   \right. \ ,
\end{equation}
\citep[e.g.,][]{babbedge06} instead of a Schechter function.  
Integrating this, then, to get the luminosity density gives
\begin{eqnarray}
\e\, j_{\e} = \int dL\ \phi(L) = 
	  \phi_*L_*\left\{ \frac{1}{2-\alpha} - \frac{1}{2-\beta} \right\}\ , 
\end{eqnarray}
assuming $\beta>2$.  Aside from direct observations, the 
UV luminosity density at high redshift can be probed with 
measurements of the photoionization rate from the Ly$\alpha$ 
forest \citep{faucher08_letter,faucher08_long}.  

\begin{figure*}
\epsscale{1.0}
\plotone{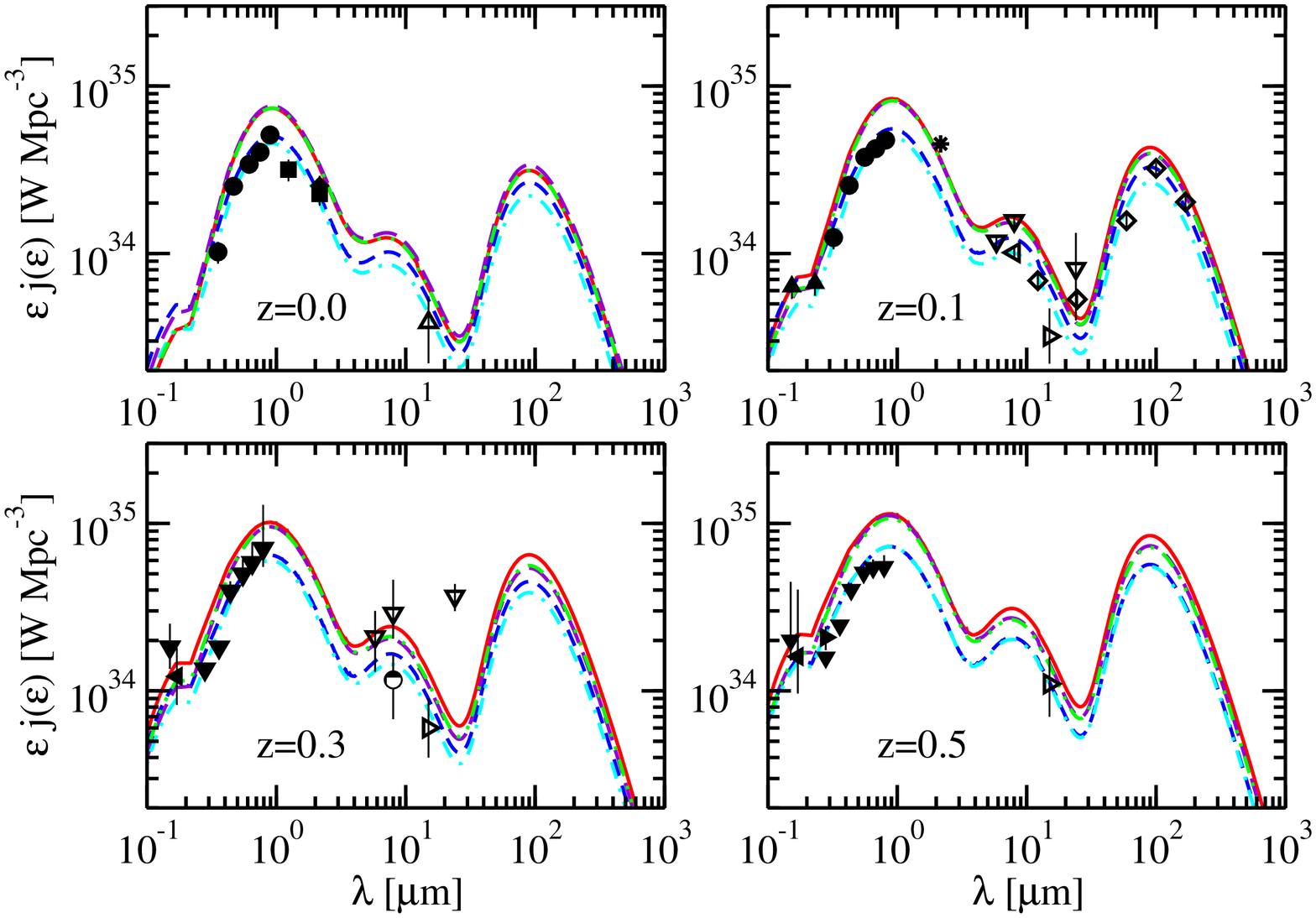}
\caption{ Data and models of the comoving luminosity density at
various redshifts.  Data are black symbols with error bars, as
follows: filled circles: \citet{blanton03}; filled squares:
\citet{cole01}; filled diamonds: \citet{kochanek01}; filled
upward-pointing triangles: \citet{budavari05}; filled
downward-pointing triangles: \citet{tresse07}; filled
leftward-pointing triangles: \citet{sawicki06}; filled
rightward-pointing triangles: \citet{dahlen07}; star: \citet{smith08};
empty upward-pointing triangles: \citet{magnelli09}; empty diamonds:
\citet{takeuchi06}; empty downward-pointing triangles:
\citet{babbedge06}; empty leftward-pointing triangles:
\citet{huang07}; empty rightward-pointing triangles:
\citet{lefloch05}; circle with cross inside: \citet{flores99}.  Curves
are our models as follows: solid red: Model A; Dashed double-dotted
green: Model B; short dashed blue: Model C; long dashed violet: Model
D; dashed-dotted cyan: Model E.
}
\label{lumdensity_lowz}
\end{figure*}

\begin{figure*}
\epsscale{1.0}
\plotone{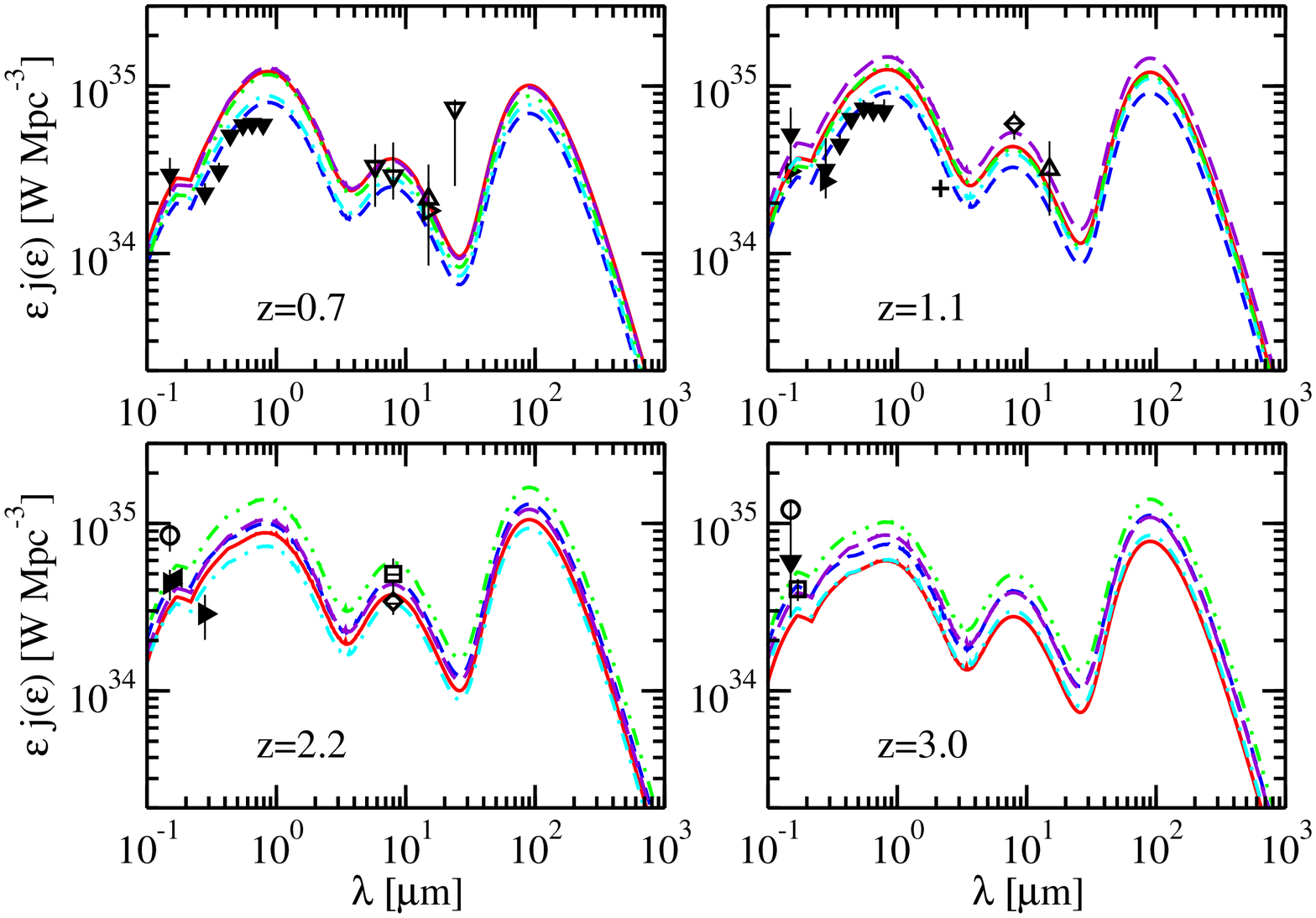}
\caption{
Data and models of the comoving luminosity density at various
redshifts.  Symbols and curves are the same as
Fig.\ \ref{lumdensity_lowz}, while additional data are from the
following sources:  
cross: \citet{cirasuolo07}; 
empty circles:  \citet{faucher08_letter,faucher08_long}; 
empty square:  \citet{reddy08};
diamonds with horizontal stripes:  \citet{caputi07}.
box with horizontal stripes:  \citet{perez05}.
}
\label{lumdensity_highz}
\end{figure*}

In terms of our theoretical formalism, the luminosity density is
dependent on the IMF, SFR, and dust parameters, and can be calculated
from eqns.\ (\ref{lumdens}) and (\ref{dust_lumdens}).  The results of
our models, plotted with measurements of the luminosity density at
various redshifts, can be seen in Figs.\ \ref{lumdensity_lowz} and
\ref{lumdensity_highz}.  Although the differing measurements do not
entirely agree, they provide a consistent picture for the
evolution of the luminosity density, particularly in the UV/optical
and at low redshift.

%\clearpage

%\clearpage

It appears that our Model C best fits the luminosity density data.
Models C and E fit the data about equally well at most wavelengths,
but model C fits the UV data at low redshift (especially $z=0.1$)
better than does model E.  These models agree with the data quite well
at low $z$, and less-well at high $z$; however, Model C still provides
a better fit to the data at high $z$ than do the other models.
Measurements at higher redshifts are less certain, and less likely to
agree with each other.  Interestingly, the luminosity densities at
high $z$ from Ly$\alpha$ measurements \citep{faucher08_long} are about
a factor of two higher than measurements from direct imaging.
\citet{faucher08_long} speculate that this may be due to a faint
population of galaxies at high redshift which have avoided detection
in deep surveys so far.  Other possible explanations for a discrepancy
at high $z$ include a changing IMF, and a changing absorption from
dust; we assume both of these are independent of redshift in our
models.  The 24 $\mu$m luminosity density from \citet{babbedge06}
increases rapidly at high $z$, and is considerably above our models
and other data.

The higher redshift luminosity density contributes little to the local
EBL.  Thus such discrepancies have little effect on the direct EBL
observations (\S\ \ref{EBLlight}) and absorption of TeV $\gamma$-rays
from nearby blazars (\S\ \ref{absorption_sec}), although it could have
implications for absorption of $\gamma$-rays from higher $z$ objects
observed with {\em Fermi} or {\em AGILE}, however, even here generally
only for the UV/optical.

\subsection{Extragalactic Background Light}
\label{EBLlight}

Direct measurement of the EBL energy density is difficult due to
contamination from foreground zodiacal and Galactic light
\citep{bernstein02, gorjian00, dwek98_obs, cambresy01, wright00,
levenson07, hauser98}.  For example, the foreground subtraction of the
optical measurements of the EBL with the {\em Hubble Space Telescope}
by \citet{bernstein02} have been vigorously debated
\citep{mattila03,bernstein05, bernstein07}.  The most accurate
IR-to-UV EBL measurements are the measurements of the far-IR with COBE
\citep{fixsen98} and BLAST \citep{marsden09}.  Galaxy counts may be
used to estimate the EBL \citep{madau00, fazio04, dole06, metcalfe03,
papovich04} but the crowded fields make this difficult, and the
unknown number of unresolved sources results in a lower limit.
\citet{levenson08} have used a Monte Carlo Markov chain analysis to
correct for crowded fields in 3.6 $\mu$m galaxy counts with the {\em
Spitzer Space Telescope}, resulting in significantly stronger lower
limits.  Their new technique, when applied to other wavelengths, could
drastically improve estimates of the EBL.  Still, the lower limits
from galaxy counts, even the improved one by \citet{levenson08}, are a
factor of a few below the direct measurements.  This indicates either
that a significant fraction of the EBL originates from extremely
faint galaxies not resolved in number counts, or there is
significantly more foreground light than assumed.
\citet{franceschini08} consider the latter to be more likely, based on
the fact that the galaxy counts are performed in fields with quite
deep exposures.

\begin{figure*}
\epsscale{0.9}
\plotone{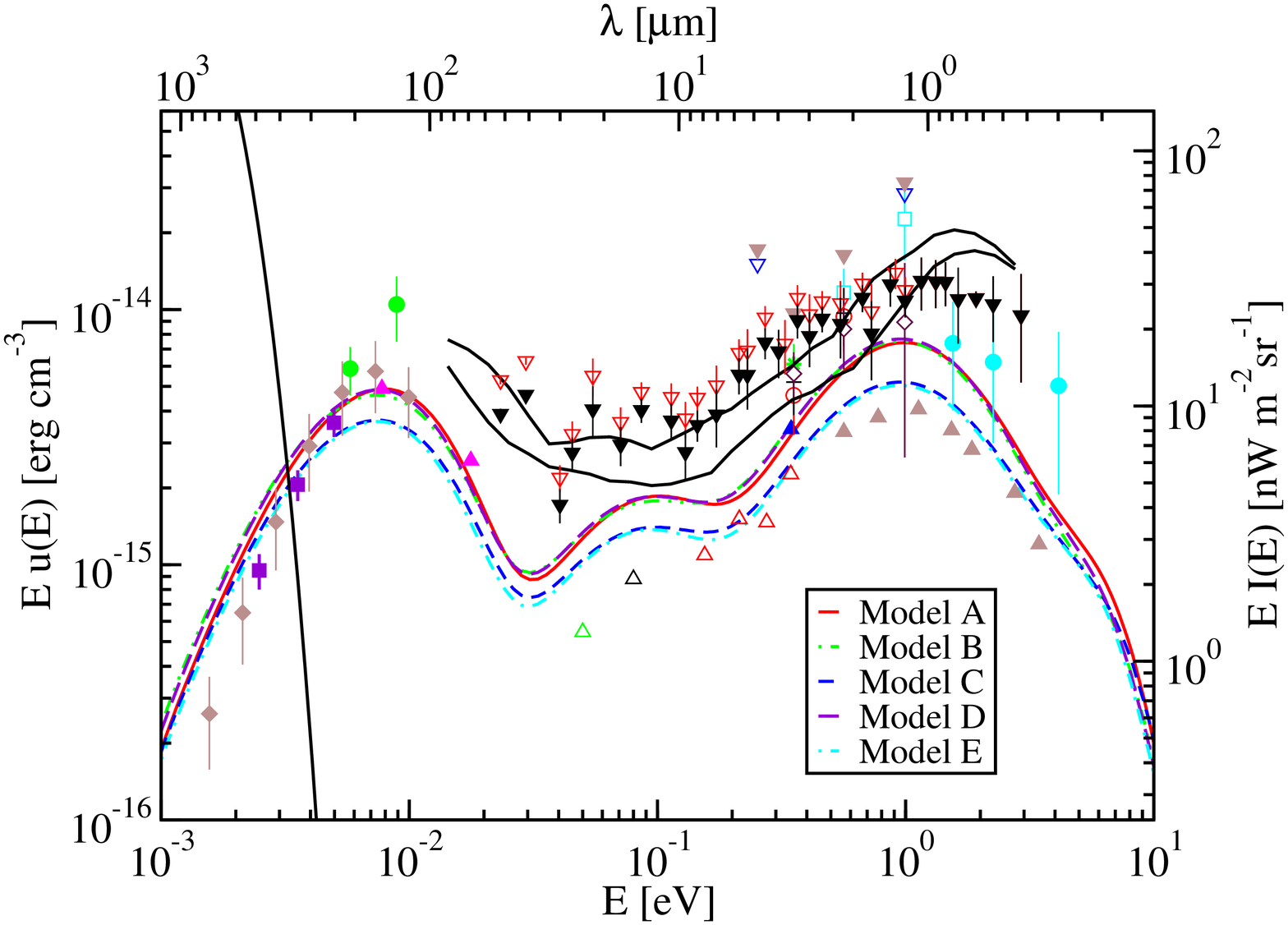}
\caption{
Our models for the EBL plotted along with measurements 
and constraints from observations.  The curves are 
our models with the same symbols as Fig.\ \ref{lumdensity_lowz}.
Measurements are from \citet[][cyan points]{bernstein02}, 
\citet[][empty red circle]{gorjian00}, 
\citet[][green asterisk]{dwek98_obs}, 
\citet[][empty cyan square]{cambresy01}, 
\citet[][black cross]{wright00}, 
\citet[][maroon diamonds]{levenson07}, 
\citet[][green filled circles]{hauser98}, 
\citet[][brown filled diamonds]{fixsen98}, 
and \citet[][violet filled squares]{marsden09}.  
Lower limits are from \citet[][red empty triangles]{fazio04}, 
\citet[][brown filled triangles]{madau00}, 
\citet[][blue filled triangle]{levenson08}, 
\citet[][magenta filled triangles]{dole06}, 
\citet[][black empty triangle]{metcalfe03}, and 
\citet[][green empty triangle]{papovich04}.  
Upper limits are from \citet[][brown filled inverted triangles]{hauser98}, 
\citet[][blue empty inverted triangles]{dwek98_obs}, 
\citet[][upper and lower black curves $\Gamma_{int}^{min}=0.67$ and 
$\Gamma_{int}^{min}=1.5$ upper limits, respectively]{mazin07}, and 
red empty and black filled inverted triangles are the  
$\Gamma_{int}^{min}=1.0$ and $\Gamma_{int}^{min}=1.5$ 
upper limits, respectively, from \citet{finke09}.  The black curve 
at long wavelengths is the cosmic microwave background.  
}
\label{EBLmodels_mine}
\end{figure*}

The EBL photon density, calculated with our models, can be seen in
Fig.\ \ref{EBLmodels_mine}, along with direct measurements, lower
limits from galaxy counts, and upper limits from TeV blazar
observations.  Models C and E are very close to the galaxy count lower
limits in the optical and UV, and in fact are below the near-IR lower
limits from \citet{levenson08} and the far-IR limits from
\citet{dole06}.  Models A, B, and D, on the other hand, seem to be
marginally above the upper limits from $\gamma$-ray observations of
blazars \citep{mazin07,finke09} although these results are not without
controversy \citep[e.g.,][]{stecker08}.  Due to how well Model C fit
the luminosity density data, we take this as our best fit model.

The integrated intensity from our models can be seen in Table
\ref{modelintegral}.  \citet{fardal07} created a simple model by
a fit which compromises between direct photometric observations and
number-count lower limits on the local EBL intensity.  They find an
integrated EBL intensity of 50 -- 130 nW m$^{-2}$ sr$^{-1}$, and a
value closer to the lower end is required to be consistent with the
$K$-band luminosity density (similar to our best fit model, model C).
\citet{nagamine06} find this value to be 43 $\pm$ 7 nW m$^{-2}$
sr$^{-1}$, which is in good agreement with our model C.

Our model of the stellar component is more accurate at low redshifts
than high redshifts, and more accurate than the dust component, since
there are more luminosity density data in this region with which to
normalize our results.  Results from our best fit model are in quite
good agreement with lower limits from galaxy counts in the near-IR
through UV, except for the recent 3.6 $\mu$m point of
\citet{levenson08}.  In the far IR, this model is slightly below the
points of \citet{dole06}, although they do fit the BLAST data
\citep{marsden09} fairly well.  The addition of a separate component
representing ultraluminous infrared galaxies, as done by
\citet{kneiske04}, may bring our model C in agreement with these other
lower limits.  However, it is unclear why this component would not be
represented in the luminosity density measurements (Figs.\
\ref{lumdensity_lowz} and \ref{lumdensity_highz}).
\citet{krennrich08} have recently used the lower limit of
\citet{levenson08} to construct their own EBL SED, which gives a
greater energy density than our model C.

\begin{figure}
\epsscale{1.0}
\plotone{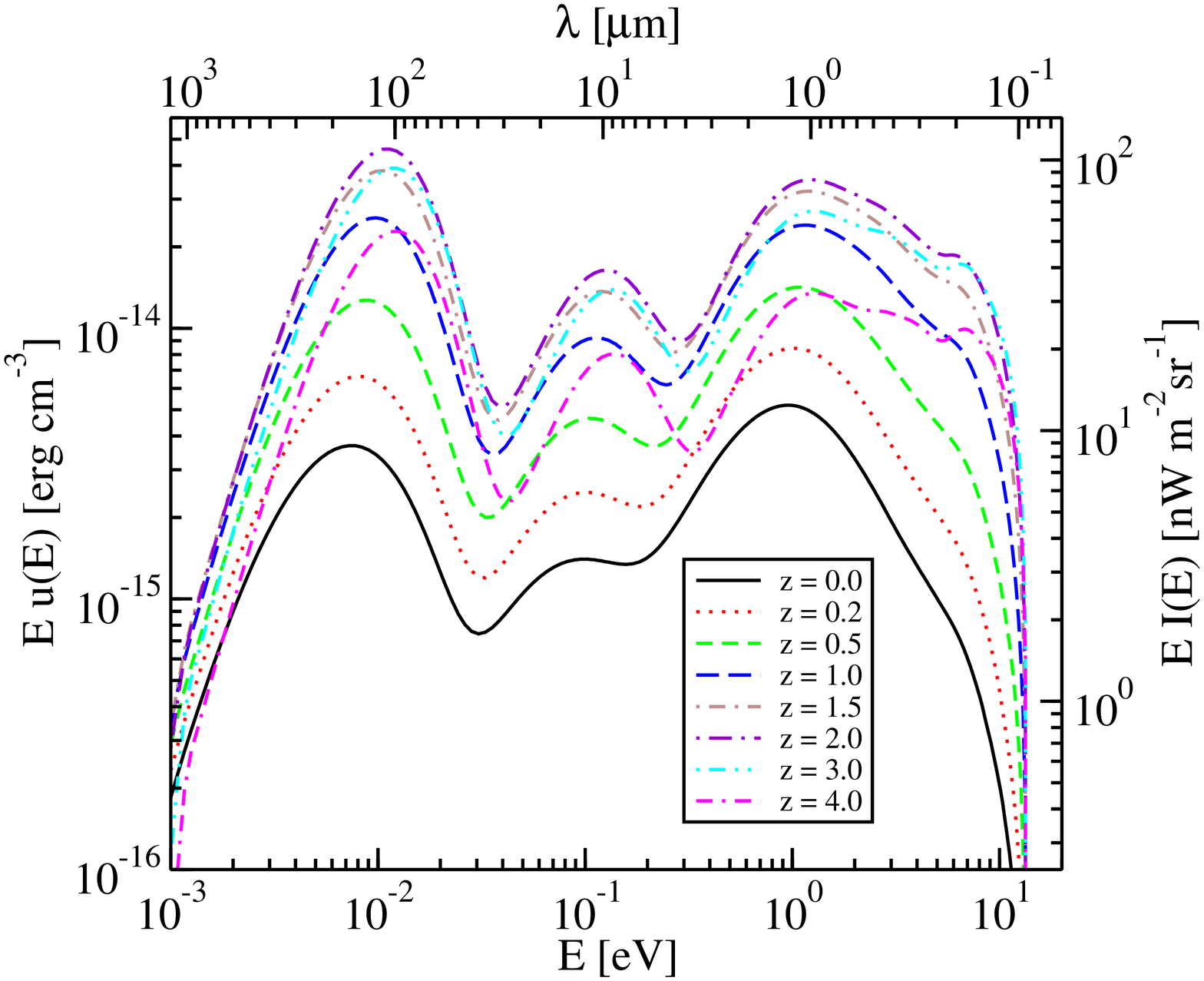}
\caption{
The proper EBL energy density as a function of proper photon 
energy for model C, for a variety of redshifts.
}
\label{timeevo_modelC}
\end{figure}

Fig.\ \ref{timeevo_modelC} shows the redshift evolution of model C.
This peaks at around $z\approx2$, which is about where the SFR peaks
(see RDF09).  Also note that at higher redshifts, there are more high
mass stars to create a larger high energy component.  The high $z$
SEDs are flatter in the optical-UV due to this relatively higher
contribution of high mass stars.  At lower $z$, the contribution is
greater from longer living low mass stars, as well as those high mass
stars which have evolved off the main sequence.  This also affects the
dust emission component.  At around $z=1$ the far-IR dust component
peaks at a greater energy density than does the optical stellar
component, and the mid-IR PAH component gets progressively greater at
higher $z$.  Since the dust absorption is greater at higher energy
(lower wavelength), where the high mass stars emit most of their
radiation, the absorption is greater at higher $z$, and thus the dust
emission is greater at higher $z$.

\begin{figure}
\epsscale{0.9}
\plotone{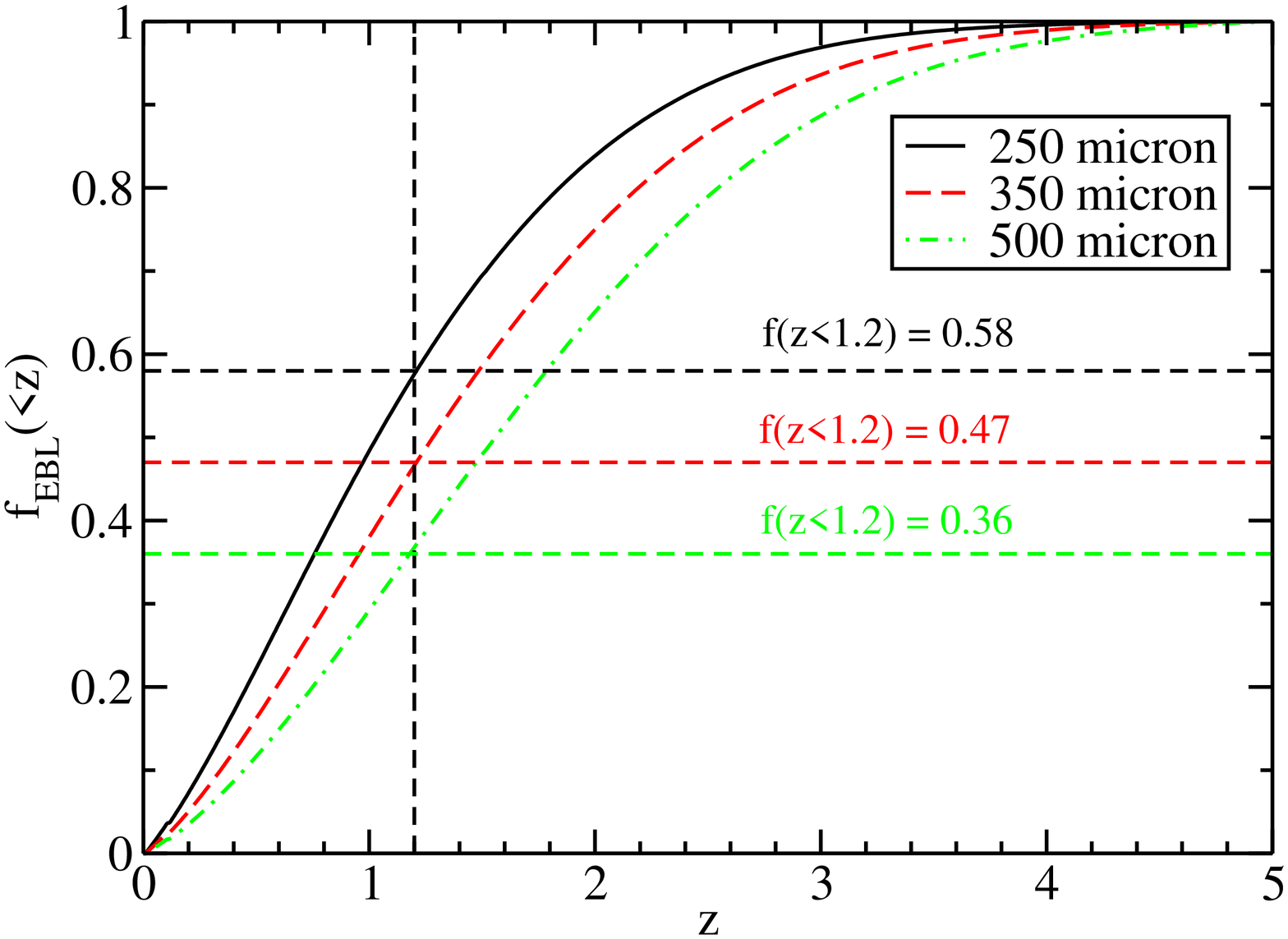}
\caption{ The fraction of the local ($z=0$) EBL which originates
below a redshift $z$ at the three BLAST wavebands \citep{marsden09} 
for our model C.
} 
\label{EBLfrac}
\end{figure}

The Balloon-borne infrared telescope BLAST found that the
fraction of galaxies with redshift $z<1.2$ which contribute to the
local EBL is $f_{EBL}(z<1.2,250\ \mu m)=0.60\pm0.11$,
$f_{EBL}(z<1.2,350\ \mu m)=0.49\pm0.10$, and $f_{EBL}(z<1.2,500\ \mu
m)=0.39\pm0.11$, where the error bars include measurement and
calibration uncertainties \citep{marsden09}.  We calculate these 
fractions for our model C, to test its consistency with these 
BLAST results, with 
\begin{eqnarray}
f_{EBL}(<z,\e) \equiv \frac{ 
	\ \int^{z}_{0}dz_1\ 
	\frac{ \ep\, j(\ep; z_1)}{(1+z_1)}\ 
	\left|\frac{dt_*}{dz_1}\right|\ 
}
{ \e\, u_{EBL}(\e;z=0)  }
\end{eqnarray}
Results from this calculation can be seen in Fig.\ \ref{EBLfrac}.  As
can be seen, these fractions are in excellent agreement with the
results from BLAST.  

%\clearpage
\begin{deluxetable}{cccc}
%\rotate
\tabletypesize{\scriptsize}
\tablecaption{
Integrated EBL intensity from our models, 
in nW m$^{-2}$ sr$^{-1}$.
}
\tablewidth{0pt}
\tablehead{
\colhead{ Model} &
\colhead{ Stellar Component } &
\colhead{ Dust Component } &
\colhead{ Total } 
}
\startdata
A & 36.0 & 25.9 & 61.9 \\
B & 36.4 & 25.5 & 61.8 \\
C & 26.5 & 20.4 & 46.8 \\
D & 37.0 & 26.1 & 63.1\\
E & 25.5 & 19.6 & 45.0\\
\enddata
\label{modelintegral}
\end{deluxetable}
%\clearpage

%\clearpage

%\clearpage

%\clearpage

\begin{figure}
\epsscale{1.0}
\plotone{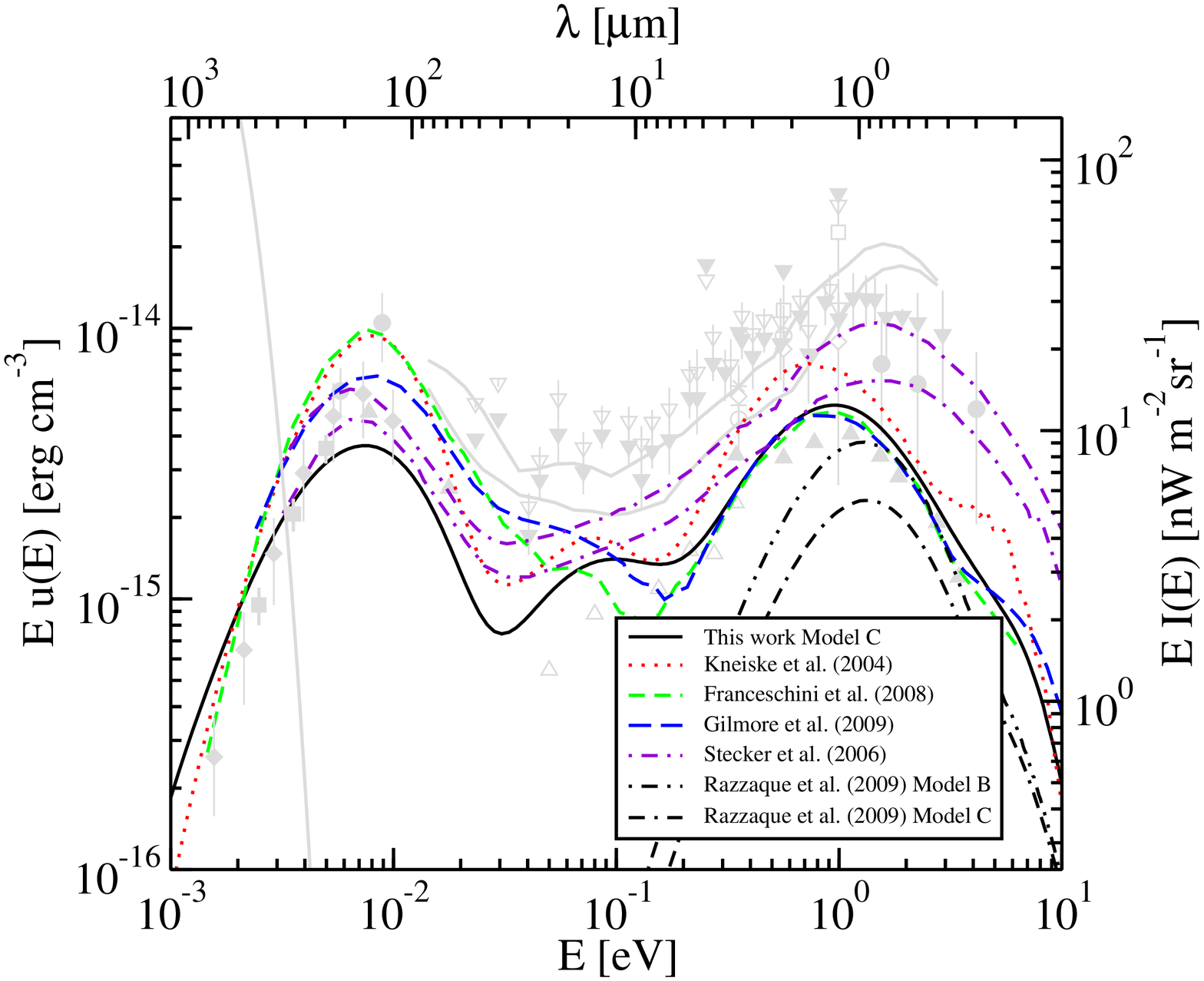}
\caption{ Our model C (solid black curve), the best fit model of
\citet[][red dotted curve]{kneiske04}, the model of \citet[][short
dashed green curve]{franceschini08}, the fiducial model of
\citet[][long dashed blue curve]{gilmore09} and the fast evolution and
baseline model from \citet[][upper and lower dot-dashed violet curves,
respectively]{stecker06}.  The double dot-dashed and the dot
double-dashed black curves are the single power-law model B and model
C from RDF09, respectively.  }
\label{EBLmodels}
\end{figure}
%\clearpage

\subsection{Comparison with other models}
\label{sec_compare}

We have plotted our best fit model, Model C, along with several other
models from the literature in Fig.\ \ref{EBLmodels}.  Note that our
model agrees remarkably well with those of \citet{franceschini08} and
\citet{gilmore09} at $\ga 0.3$ eV ($\la 5$ $\mu$m).  Of the models
presented in Fig.\ \ref{EBLmodels}, only the best fit model of
\citet{kneiske04} and the fast evolution model of \citet{stecker06}
agree with the recent 3.6 $\mu$m lower limit from \citet{levenson08}.  

Two previous models from RDF09 are plotted here, using the single
power-law version of the stellar luminosity-mass relation.  These
previous versions of our model include the contribution to the EBL
from main-sequence stars.  Model B from RDF09 was considered our best
fit model from this previous work, while we have plotted the old Model
C for comparison with our best fit model from this work.  It is clear
that post-main sequence stars have a considerable effect on the
result, as the C models differ by as much as an order of magntitude at
longer wavelengths ($\sim 1-10\ \mu m$).  The model in this work is
more intense longward of $\sim 1\ \mu$m due mainly to giant stars,
while shortward of $\sim 0.3\ \mu$m the inclusion of white dwarfs
explains the discrepancy between these works.  

The modeling technique of \citet{kneiske02,kneiske04} is quite similar
to ours: they also integrate over stellar parameters and treat dust
extinction and re-emission in a similar fashion.  Our results,
however, differ significantly from theirs due to our use of updated
parameters for the SFR, IMF and dust extinction.  The best fit model
of \citet{kneiske04} also includes separate components from 
ultraluminous infrared galaxies (ULIGs) and normal galaxies, which,
if applied to our model (as mentioned above), could result in a local
EBL energy densities more in agreement with the lower limits of
\citet{dole06} and \citet{levenson08}, with which the
\citet{kneiske04} best fit model is in agreement.  Their infrared
component, however, considerable overpredicts the $250\ \mu$m
measurements from BLAST \citep{marsden09} and COBE \citep{fixsen98}.
Disentangling the dust extinction from these two different
populations, using the realistic \citet{driver08} \citep[or
the][]{popescu09} model would be considerably more complicated.
Furthermore, greater dust emission may contradict the luminosity
density data at low $z$ (Fig.\ \ref{lumdensity_lowz}).

\citet{stecker06} have used a backwards evolution model based on
observations of the infrared luminosity function to calculate the IR
EBL energy density.  They then use a forward evolution model similar
to \citet{salamon98} to calculate the optical/UV EBL, and normalize it
to their backward-evolution calculation.  Our model peaks in the IR at
an intensity relatively close to their baseline model, both of which
are below the IR lower limits of \citet{dole06}.  However, our models
differ by as much as a factor of 8 in the UV.  This could be
because extrapolating below 10 $\mu$m may not be appropriate for
determining the infrared background \citep{lagache03}.  This
wavelength is critically important to the models of \citet{stecker06}
because this is where they normalize the results of their
forward-evolution modeling, and the optical/UV region is the most
important for $\g$-ray absorption (\S\ \ref{absorption_sec}).  
Furthermore, their model has neglected dust extinction, which has a
considerable effect at UV wavelengths.  We note that the far IR peak
of their baseline model, which is based in IR luminosity density data,
is within 10\% of our far-IR peak, which is fit to similar IR
luminosity density data.

\citet{gilmore09} have presented models based on semi-analytic
models of galaxy evolution, using cosmological parameters from WMAP
and normalizing their semi-analytic model's luminosity density
prediction to local luminosity density measurements.  They also
include a component from quasars and realistically treat extinction of
ionizing radiation by the intergalactic medium, both of which have
implications for the UV portion of the EBL at high redshift.
\citet{gilmore09} only treat the UV-optical portion of the EBL, while
their further work will be presented in Gilmore et al. (in
preparation).  This is an update of previous, similar models
\citep{primack99,primack01,primack05}.  The stellar component of our
model probably agrees with the \citet{gilmore09} fiducial model so
well because we have both made sure our models agree well with
luminosity density measurements (\S\ \ref{lumdens_section}).  However,
at lower energies (longer wavelengths) their model intensity is 
as much as double ours.  Moreover, their predicted intensity is above
the $250\ \mu$m BLAST measurement \citep{marsden09}, although it is
consistent with the COBE results \citep{fixsen98}.  

The model of \citet{franceschini08} is based on integrating luminosity
functions at various wavelengths and redshifts to generate an EBL
energy density.  Again, their model agrees well with ours at higher
energies but at $20\ \mu$m they predict about three times the
intensity as we do.  The infrared component of their model is
comparable to that of the best fit model of \citet{kneiske04}, and
similarly, it over-predicts the $250\ \mu$m BLAST \citep{marsden09}
measurement and the COBE \citep{fixsen98} infrared data.  

\section{Absorption of $\gamma$-rays}
\label{absorption_sec}

Once the energy density of the EBL has been calculated using eqn.\
(\ref{n_EBL}), the absorption optical depth of $\g$-ray photons as a
function of observed $\g$-ray photon energy, $\e_1$, can be calculated
by
\begin{eqnarray}
\label{taugg}
\tau_{\g\g}(\e_1, z) = \frac{c\pi r_e^2}{\e_1^2 m_ec^2}\ 
\int^z_0 \frac{dz^\prime}{(1+z^\prime)^2}\ 
\left| \frac{dt_*}{dz^\prime}\right|\ 
\nonumber \\ \times
\int^\infty_{\frac{1}{\e_1(1+z^\prime)}} d\e_{p} 
\frac{ \e_{p}u_{EBL,p}(\e_{p};z^\prime)}
{\e_p^{4}}\ 
\bar{\phi}(\e_{p}\e_1(1+z^\prime))\ ,
\end{eqnarray}
where 
$|dt_*/dz|$ is given by eqn. (\ref{dtdz}), 
$\e^{p}$ 
is the proper frame EBL photon energy, 
$\e^{p}\,u^{p}_{EBL}(\e^{p};z^\prime)$  
is the proper frame EBL energy density given by eqn. (\ref{n_EBL}), 
\begin{eqnarray}
\bar{\phi}(s_0) = \frac{1+\beta_0^2}{1-\beta_0^2}\ln w_0 - 
\beta_0^2\ln w_0 - \frac{4\beta_0}{1 - \beta_0^2}
\\ \nonumber 
+ 2\beta_0 + 4\ln w_0 \ln(1+w_0) - 4L(w_0)\ ,
\end{eqnarray}
$\beta_0^2 = 1 - 1/s_0$, $w_0=(1+\beta_0)/(1-\beta_0)$, and 
\begin{eqnarray}
L(w_0) = \int^{w_0}_1 dw\ w^{-1}\ln(1+w)\ 
\end{eqnarray}
\citep{gould67,brown73}.  

\begin{figure*}
\epsscale{1.0}
\plotone{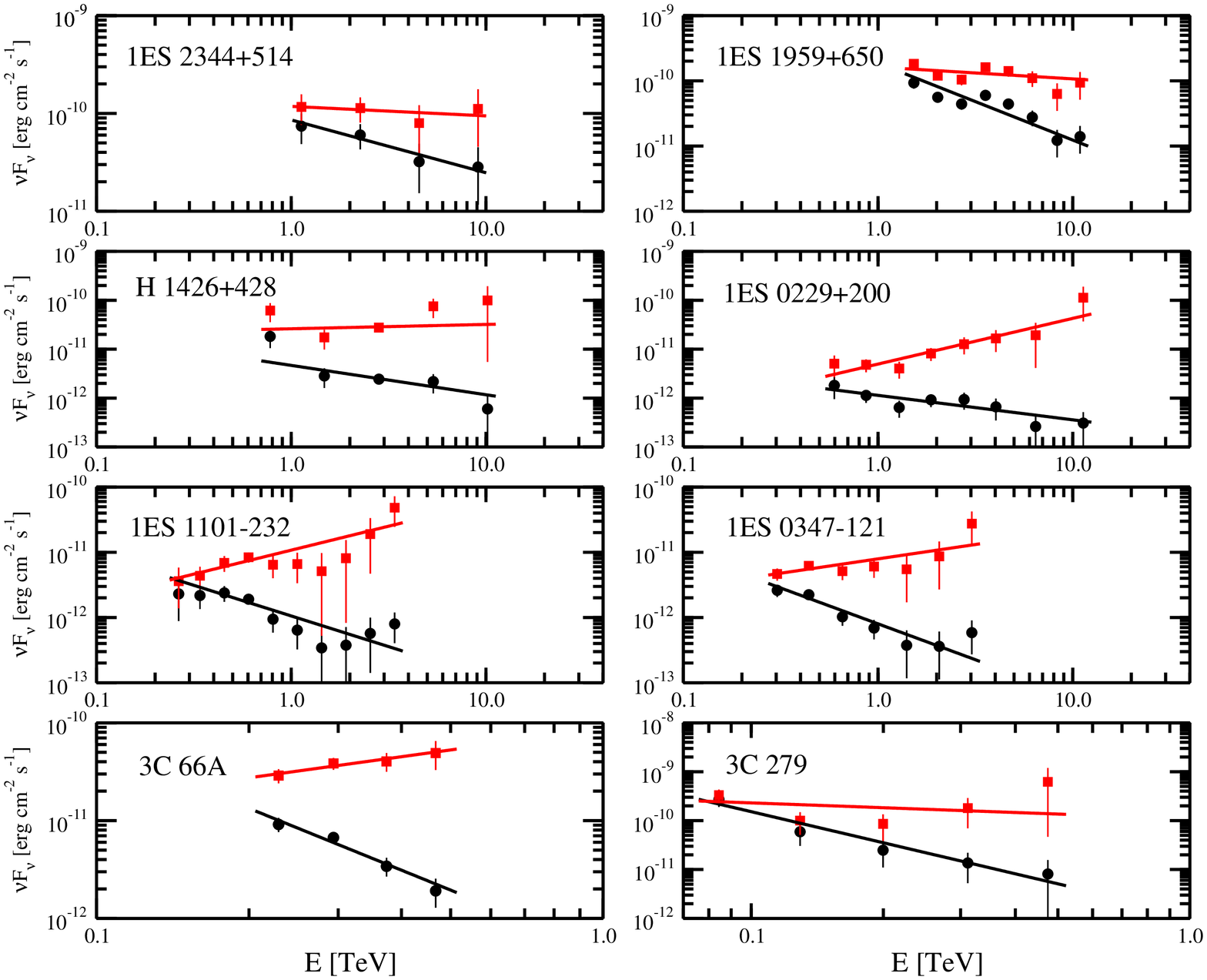}
\caption{
TeV blazar spectra observed (circles) and deabsorbed by our model C
(squares).  
}
\label{taudeabsorb}
\end{figure*}

We have used our model C to deabsorb several of the blazars presented
in \citet{finke09}.  We choose blazars that have a low $\xi$ 
parameter, defined in \citet{finke09} to be 
\begin{eqnarray}
\xi = \left(\frac{TeV}{E_{max}}\right)
	\frac{\Gamma_{obs}-\Gamma_{int}^{min}}{z}
	\ln\left(\frac{E_{max}}{E_{min}}\right)\ ,
\end{eqnarray}
(not to be confused with $\xi(m)$, the IMF in \S\ \ref{starlight})
where $E_{min}$ and $E_{max}$ are the minimum and maximum energies of
the VHE $\g$-ray spectrum, $\Gamma_{obs}$ is the observed photon
spectral index, and $\Gamma^{min}_{int}$ is an assumed minimum
intrinsic photon spectral index, for which we use $\Gamma_{int}=1.5$,
the lowest value one would expect from na\"ive test particle shock
acceleration theory where particles are accelerated with number
indices steeper than $-2$ \citep[for a review, see][]{blandford87}
although see \citet{stecker07_accel} for a different interpretation.
The lower the parameter $\xi$, the lower the deabsorbed spectral index
$\Gamma$ should be.  We also choose the VHE $\g$-ray spectra of two
bright, distant quasars, \object{3C 66A} and \object{3C 279} to
deabsorb, to give a greater range in redshift.  The deabsorbed spectra
are fit with power-laws, and the results can be seen in Fig.\
\ref{taudeabsorb} and Table \ref{blazarabstable}.  Note that in
Fig.\ \ref{taudeabsorb} the $\nu F_{\nu}$ spectra is plotted, where
$\Gamma=1.5$ corresponds to $\nu F_{\nu}\propto E^{0.5}$.  The results
are generally consistent with a minimum intrinsic $\Gamma_{int}$ of
1.5, as this value is within the error bars for all of the sources.

%\clearpage

%\clearpage
\begin{deluxetable*}{lccccl}
%\rotate
\tabletypesize{\scriptsize}
\tablecaption{
Deabsorbed TeV Blazars
}
\tablewidth{0pt}
\tablehead{
\colhead{ Blazar } &
\colhead{ Redshift } &
\colhead{ Observed $\Gamma$ } &
\colhead{ Deabsorbed $\Gamma$ } &
\colhead{ $\xi$ } &
\colhead{ Reference } 
}
\startdata
\object{1ES 2344+514} & 0.044 & $2.54\pm0.18$ & $2.10\pm0.68$ & 4.7 & \citet{schroedter05_2344} \\
\object{1ES 1959+650} & 0.047 & $3.18\pm0.17$ & $2.18\pm0.33$ & 4.7 & \citet{aharonian03_1959} \\
\object{1ES 1426+428} & 0.129 & $2.60\pm0.60$ & $1.91\pm0.58$ & 2.2 & \citet{aharonian03_1426} \\
\object{1ES 0229+200} & 0.139 & $3.09\pm0.26$ & $1.07\pm0.45$ & 1.8 & \citet{aharonian07_0229} \\
\object{1ES 1101-232} & 0.186 & $2.94\pm0.20$ & $1.27\pm0.46$ & 5.7 & \citet{aharonian07_1101} \\
\object{1ES 0347-121} & 0.188 & $3.10\pm0.25$ & $1.56\pm0.43$ & 6.5 & \citet{aharonian07_0347} \\
\object{3C 66A} & 0.444\footnotemark[1] & $4.10\pm0.72$ & $1.28\pm0.98$ & 19.6 & \citet{acciari09_3c66a} \\
\object{3C 279} & 0.536 & $4.11\pm0.68$ & $2.33\pm0.89$ & 16.0 & \citet{albert08_3c279} \\
\enddata
\label{blazarabstable}
%\footnotemark[1]{Although we use 0.444 as the redshift of 3C 66A it 
%is not well-known; see \citet{finke08_redshift}.}
\end{deluxetable*}
%\clearpage

\begin{figure}
\epsscale{1.0}
\plotone{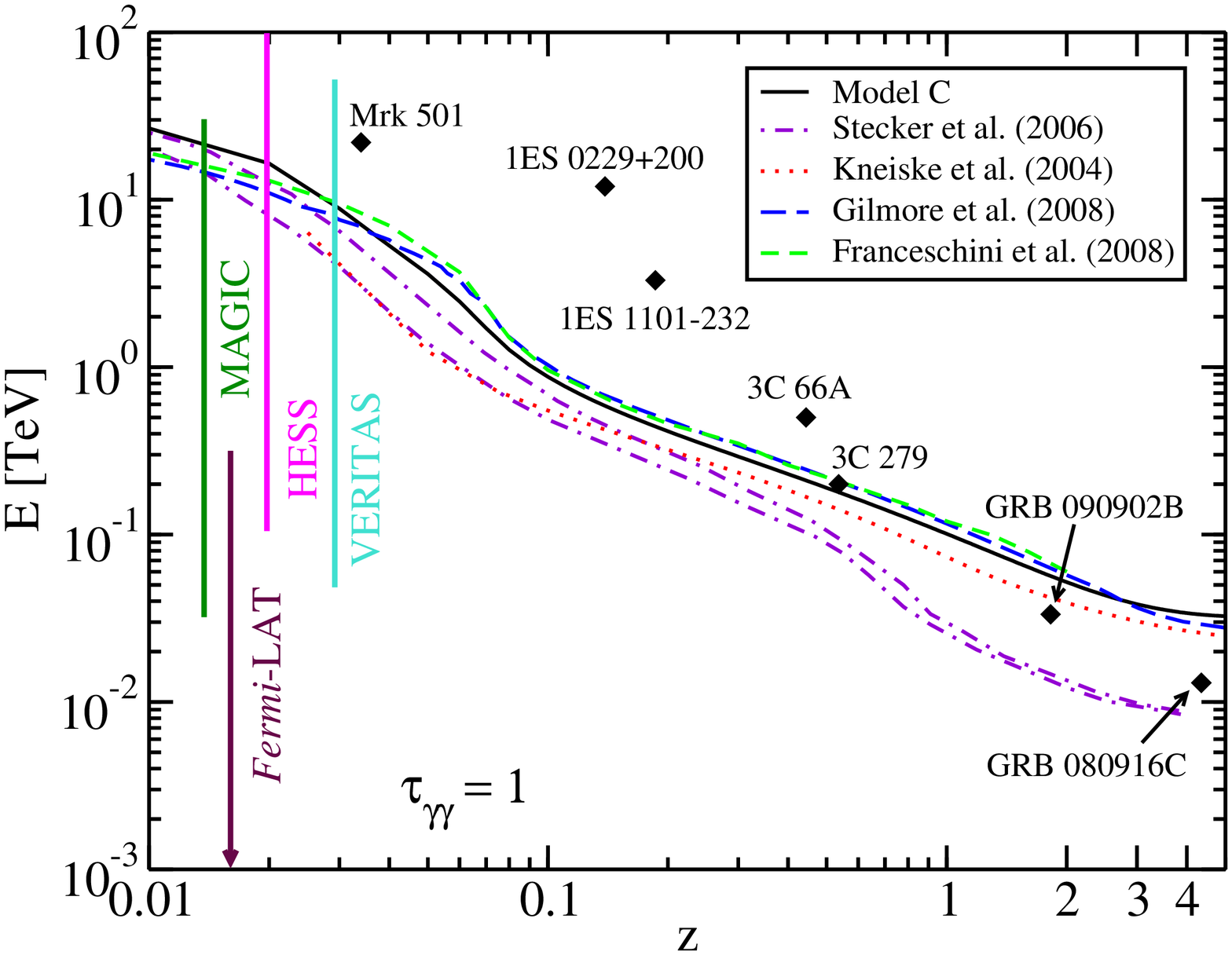}
\caption{ 
A plot of $\tau_{\g\g}=1$ for several EBL models including our model
C, as a function of redshift.  Curves signify the same models as in
Fig.\ \ref{EBLmodels}.  Also plotted by the filled diamonds are the
maximum $\g$-ray photon energy bins from blazars observed with
atmospheric Cherenkov telescopes and the GRBs 080916C and
090902B observed with the {\em Fermi}-LAT.
}
\label{faziostecker}
\end{figure}

A plot of the energy at which the universe becomes optically thick to
$\g$-rays, defined where $\tau_{\g\g}=1$ is given in Fig.\
\ref{faziostecker}, which is often known as the Fazio-Stecker relation
\citep{fazio70}, for several models including our Model C.  Also
plotted are the maximum photon energy bin of several blazars, observed
with atmospheric Cherenkov telescopes \citep[see][for a list and
references]{finke09} and the GRBs 080916C \citep{abdo09_080916c}
and 090902B \citep{abdo09_090902b} observed with the {\em Fermi}-LAT.
The VHE $\g$-rays from many blazars are highly attenuated by the EBL,
since several are considerably above the $\tau_{\g\g}=1$ for all
models, and the highest energy photons from GRBs constrain the
EBL at high redshifts.  Also note that the universe will be optically
thin to 20 GeV and lower photons over all redshifts for all models
except those of \citet{stecker06}.

\begin{figure*}
\epsscale{1.0}
\plotone{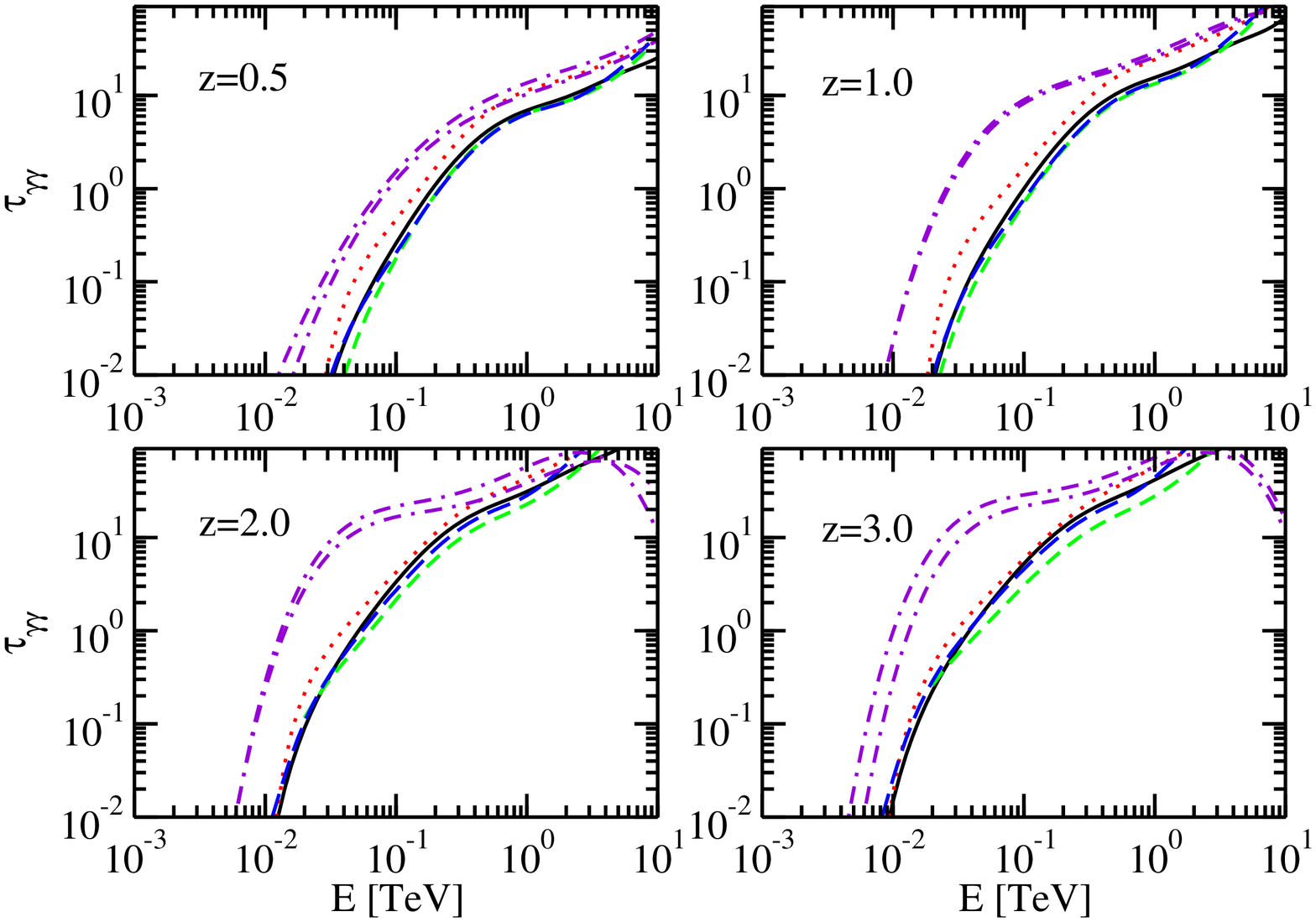}
\caption{
Absorption optical depth calculation for several EBL models.  
Curves signify the same models as in Fig.\ \ref{EBLmodels}.  
}
\label{opacity_compare}
\end{figure*}

We have plotted the absorption optical depth for our model and several
others in Fig.\ \ref{opacity_compare}.  The models are optically thin
($\tau_{\g\g}<1$) at 200 GeV for $z<3$, except the models of
\citet{stecker06}.  Observing a large number of high-energy photons
with observatories such as {\em Fermi} and {\em AGILE} from high-$z$
sources could rule out these models.  Our model is moderately more
opaque than those of \citet{gilmore09} and \citet{franceschini08} at
these redshifts, although it would be difficult to distinguish these
models based on $\g$-ray observations.

%\clearpage

%\clearpage
%\clearpage

\section{Summary and Conclusions}
\label{summary}

We have developed a model for the UV through far-IR EBL from
direct stellar radiation and stellar radiation which is absorbed and
reradiated by dust. This approach extends RDF09, which applied between
$\sim 1$--10 eV.  Our best fit model, Model C, is consistent
with the collection of measures of SFR by \citet{hopkins06}, and a
variety of luminosity density data at various redshifts $z\le 3$.  It
does not require complex stellar structure
codes or semi-analytic models of galaxy formation.  Our model is most
accurate at low-$z$ and high $\e$ (short wavelengths), which is the
energy and redshift range that is useful for calculating the
absorption of TeV $\g$-rays from nearby blazars by electron-positron
pair production.  The EBL energy density of Model C is generally
consistent with lower-limits from galaxy counts and is below direct
EBL measurements except for the lower limits at 3.6 $\mu$m
\citep{levenson08} and 60 $\mu$m \citep{dole06}.  It also agrees well
with the recent EBL models of \citet{gilmore08} and
\citet{franceschini08} in the near-IR through UV, although less well
at longer wavelengths.  Our Model C is significantly below the
models of \citet{kneiske04} and \citet{stecker06} for all wavelengths.

The SFR, IMF, and dust parameters in our best fit EBL model were
chosen to agree well with luminosity density measurements,
particularly those at low-$z$ where the data is best, and to give
precedence to luminosity density measurements over local EBL
measurements when deciding the quality of an EBL model.  Although
different luminosity density measurements are not always consistent
with each other, as one can see in Figs.\ \ref{lumdensity_lowz} and
\ref{lumdensity_highz}, they are less controversial than measurements
of the local EBL intensity.  For example, \citet{madau00} claim that
their galaxy count measurements resolve almost all of the flux in the
EBL.  \citet{bernstein02} suggest, however, that those measurements
fail to take into account the faint portions of the distant galaxies.
The more direct measurements of \citet{bernstein02} give higher values
than galaxy counts, but their measurements have also been criticized
for the subtraction of zodiacal foreground light
\citep[e.g.][]{mattila03,bernstein05,bernstein07}.  To reconcile these
differing inferences, \citet{totani01} suggest the possibility that
this discrepancy can be resolved by a faint EBL component outside of
normal galaxy populations.

By constructing an EBL model consistent with luminosity-density
observations, we have found that an EBL energy density very close to
the lower limits from number counts is required.  Similar conclusions
were reached by \citet{fardal07}.  They point out that their results
are strongly dependent on the IMF, which may vary over the history of
the universe.  Although radiation hydrodynamic simulations indicate
that radiative feedback could lead to a universal IMF in the local
universe \citep{bate09}, a detailed modeling of a large amount of
Sloan Digital Sky Survey data indicates that fainter galaxies
generally have an IMF which produces fewer massive stars than brighter
galaxies \citep{hoversten08}.  This result has further evidence in the
ratio of H$\alpha$ to far-UV flux in \ion{H}{1}-selected galaxies
\citep{meurer09}.  \citet{babbedge06} suggested that a non-universal
IMF may be responsible for their measured rapid increase in the 24
$\mu$m luminosity density (\S\ \ref{lumdens_section}).

\citet{georganopoulos08} have suggested a new method for measuring the
EBL.  They point out that Compton scattering of the infrared component
of the EBL by the radio lobes of Fornax A could be detectable by {\em
Fermi} in $\sim 2$ years in scanning mode, and that it may also be
possible to put upper limits on the optical EBL component in the
absence of {\em Fermi} detection.  This will be particularly important
for distinguishing our best fit model from those which predict
significantly higher IR emission, such as those of \citet{kneiske04}
or \citet{franceschini08}.

In the near future, surveys such as the Dark Energy Survey, and the
Sloan Digital Sky Survey II will detect $\sim 10^3$ core-collapse
supernovae per year, and Pan-STARRS and the Large Synoptic Survey
Telescope will detect $\sim 5 \times 10^5$ core-collapse supernovae
per year \citep{lien09}.  This will lead to extremely precise
measurement of the SFR, as the core collapse supernovae rate is a good
tracer of the high mass SFR, although modeling dust attenuation will
still be a significant challenge.  This in turn will lead to strong
constraints on models of the EBL and luminosity density.  Detection of
the neutrino background from core-collapse supernovae will further
constrain the SFR and EBL \citep[e.g.,][]{horiuchi08}. Combinations of
these various measurements mean that very soon the local EBL will be
known to much greater precision.

We have used our best fit EBL model (Model C) to calculate the
absorption optical depth to $\g$-rays from cosmological sources, valid
over all relevant $\g$-ray energy ranges and redshifts.  Our
results show that absorption is quite significant for blazars observed
at TeV energies by atmospheric Cherenkov telescopes, and that
de-absorbed VHE blazar spectra give results generally in agreement
with particle acceleration theory.  We also find that the universe is
transparent ($\tau_{\g\g}\la 1$) for all redshifts at energies less
than 20 GeV, which are those most relevant to the {\em
Fermi}-LAT.\footnote{The results of our EBL energy density and
opacity calculations for use in the study of extragalactic $\g$-ray
sources are available in electronic form by requests to the authors.}.

\acknowledgements 

We are grateful to the anonymous referee for useful comments which
have improved this work, including the discovery of an error in
the original version of Fig.\ 5.  We thank P. Eggleton and C. Tout for
correspondence regarding corrections to their stellar formulae, and
A. Franceschini, R. Gilmore, and T. Kneiske for correspondence related
to their EBL models. This work was supported by the Office of Naval
Research and GLAST Science Investigation DPR-S-1563-Y, and by NASA
Swift Guest Investigator Grant DPR-NNG05ED41I.

%!****************************************************

%\begin{thebibliography}{}

\bibliographystyle{apj}
\bibliography{references,blazar_ref,EBL_ref}
%\end{thebibliography}

\end{document}